\documentclass[10pt, a4paper, twocolumn]{article} 
\usepackage[colorlinks=true,linkcolor=black,anchorcolor=black,citecolor=black,filecolor=black,menucolor=black,runcolor=black,urlcolor=blue]{hyperref}
\usepackage[normalem]{ulem}

\usepackage[english]{babel} 
\usepackage{microtype} 
\usepackage{amsmath,amsfonts,amsthm} 

\usepackage[svgnames,table,xcdraw]{xcolor} 

\usepackage[hang, small, labelfont=bf, up, textfont=it]{caption} 

\usepackage{booktabs} 

\usepackage{lastpage} 

\usepackage{graphicx} 
\usepackage{xurl}

\usepackage{enumitem} 
\setlist{noitemsep} 

\usepackage{sectsty} 
\allsectionsfont{\usefont{OT1}{phv}{b}{n}} 

\usepackage{geometry} 

\usepackage[T1]{fontenc} 
\usepackage[utf8]{inputenc} 

\usepackage{XCharter} 

\usepackage{fancyhdr} 
\fancypagestyle{firstpage}{ 
	\fancyhf{}
}

\newcommand{\authorstyle}[1]{{\large\usefont{OT1}{phv}{b}{n}\color{DarkBlue}#1}} 

\newcommand{\institution}[1]{{\footnotesize\usefont{OT1}{phv}{m}{sl}\color{Black}#1}} 

\usepackage{titling} 

\newcommand{\HorRule}{\color{DarkGoldenrod}\rule{\linewidth}{1pt}} 

\pretitle{
	\vspace{-30pt} 
	\HorRule\vspace{10pt} 
	\fontsize{28}{36}\usefont{OT1}{phv}{b}{n}\selectfont 
	\color{DarkBlue} 
}

\posttitle{\par\vskip 15pt} 

\preauthor{} 

\postauthor{ 
	\vspace{10pt} 
	\par\HorRule 
	\vspace{20pt} 
}

\usepackage{lettrine} 
\usepackage{fix-cm}	

\newcommand{\initial}[1]{ 
	\lettrine[lines=3,findent=4pt,nindent=0pt]{
		\color{DarkGoldenrod}
		{#1}
	}{}%
}

\usepackage{xstring} 

\newcommand{\lettrineabstract}[1]{
	\StrLeft{#1}{1}[\firstletter] 
	\initial{\firstletter}\textbf{\StrGobbleLeft{#1}{1}} 
}

\usepackage[autostyle=true]{csquotes} 

\title{Python4Physics: \\ A physics outreach program} 

\author{
	\authorstyle{R.~A.~Brice\~no\textsuperscript{1}} 
	\newline\newline 
	\textsuperscript{1}\institution{Department of Physics, University of California, Berkeley, CA 94720, USA \\
 Nuclear Science Division, Lawrence Berkeley National Laboratory, Berkeley, CA 94720, USA}\\ 
	\text{Email: rbriceno@berkeley.edu} \\ \text{\href{https://orcid.org/0000-0003-1109-1473}{ORCID: 0000-0003-1109-1473}}
 \\
 \\
 	\authorstyle{T.~C.~Rogers\textsuperscript{2,3}} 
	\newline\newline 
	\textsuperscript{2}\institution{Department of Physics, Old Dominion University, Norfolk, VA 23529, USA}\\ 
    \textsuperscript{3}\institution{Jefferson Lab, 12000 Jefferson Avenue, Newport News, VA 23606, USA} \\
	\text{Email: trogers@odu.edu} \\ \text{\href{https://orcid.org/0000-0002-0762-0275}{ORCID: 0000-0002-0762-0275}}
}

\date{\today} 

\begin{document}

\maketitle 

\thispagestyle{firstpage} 


\lettrineabstract{We describe a summer outreach 
program developed to cultivate interest in physics in particular and physical sciences more broadly among high school and early college students using small projects in the Python programming language. We discuss the lessons we learned in the hopes that they will be valuable to other physicists in planning their own outreach efforts. We also provide links to resources and materials from the Python4Physics program, which we hope might be useful in other outreach programs.}

\section{Outreach to high schools}
\label{s.outreach}

Professional research scientists frequently express a desire to cultivate a greater interest in science in a wider population of young people. However, a university or lab environment provides a limited set of opportunities to interact with high school youth (and other prospective scientists) who are truly in their most formative years. It is often during high school years that many scientists are initially inspired to follow a path toward a career in science, and it is clearly a missed opportunity that scientists rarely interact directly with students at this stage. To address this, for the last five years, we have been actively reaching out to local high schools, both domestically and abroad, to cultivate greater interest in Science, Technology, Engineering, and Math (STEM). To increase the likelihood that those high school students will choose and stay in a STEM career, we created a free course, called Python4Physics,  to teach students the minimal scientific background needed to perform computing-based research in a fun and accessible way. 

Our purpose with this document is to provide other researchers hoping to develop their own broad-based outreach programs with resources and an account of lessons learned from our own experiences.  

\section{Python4Physics}
\label{s.p4p}

Our first Python4Physics course took place in person at Old Dominion University during the summer of 2019. The basic structure of the course was organized around lessons in the Python programming language. A small number of introductory slides were presented, and then students gathered in groups to solve fun but simple and open-ended problems on the computer. Examples of some of the problems were projectile motion projects with graphing and animation, elastic collisions in both nonrelativistic and relativistic scenarios, and the fitting of a function in a manner akin to the identification of a particle discovery in particle physics. All lectures were designed assuming the students had not yet seen calculus, but we did assume some experience with basic algebra. 

By structuring the course around programming-based tasks, we aimed to initiate students into scientific computer programming more generally, thereby equipping them with tools that would be helpful in a wide range of potential future scientific paths. At the same time, by emphasizing specific scientific problems we hoped to give the students a chance to explore generally scientific ideas in an open-ended way in the style of a practicing professional scientist. This avoided having the course become simply another ``programming class.'' We used the Python programming language because it is a popular high-level language with a vast amount of free online resources and support. Our guiding principle has been to maximize the motivational and inspirational aspects of these projects rather than technical details.

In the summer of 2019, these meetings took place twice a week and lasted for several hours. After a short introductory explanation of the task or project, our role was simply to move around the room assisting student groups as they worked. In this, we had help as well from volunteer undergraduate students, graduate students, and postdoctoral researchers. In this way, the program also provided learning opportunities for our early career researchers to develop their own teaching, outreach, and communication skills.

In order to reduce our burden and the pressure on the students, we decided not to provide grades on the projects. Instead, we gave them verbal feedback, which was largely positive reaffirmation and constructive criticism. We also provided them with our code that includes solutions to the projects. We encouraged them to only look at our code \emph{after} they had made a strong independent attempt to solve the problem. 

In the summer of 2020, the global pandemic forced us to reorganize the program around online activities, conducted via \href{https://www.youtube.com/watch?v=feaLk9zxAEM&list=PLVs7RfracJBZOHaarbbn0BP17hoxvq2Pi}{YouTube},  and this impacted certain aspects of the course itself. A positive outcome was that it allowed us to expand the reach of the program to a much larger~(100-300 students per year), and more international audience~(30-70 countries represented per year). We also expanded the scope of topics that we discussed to include more general scientific and computational topics outside of physics, including topics motivated by current events such as the basics of modeling epidemics. The popularity of Python4Physics in Latin American was sufficiently large, that with the help of postdocs and graduate students, in the summer of 2021, we hosted a Spanish version of the class, called \href{https://www.youtube.com/watch?v=Tb3S8vB4mmg&list=PLVs7RfracJBbj3BnHqaSOqwy1DFRVZFGL}{Python y Ciencias}.\footnote{We were able to host the program remotely and reach large international audiences thanks to the help of the \href{https://physics.berkeley.edu/visiting-students/reyes-remote-experience-young-engineers-and-scientists}{Remote Experience for Young Engineers and Scientists (REYES)} program.}  

The largest drawback of hosting these classes remotely was the lack of personal interaction with the students. It was also challenging to imitate the hands-on environment that we had in person. This is motivating us and others to return to an in-person, or at least hybrid, environment.\footnote{For more information about the hybrid Python4Physics class being hosted in the summer of 2024 by the University of California, Berkeley Physics Department, including the format of the \href{https://docs.google.com/forms/d/e/1FAIpQLSc6KvknrKx80AP1bggk34KyCrCTEpRHMBhsn9VNQYYjRzmCWA/viewform}{registration page}, go the  \href{https://physics.berkeley.edu/visiting-students/python4physics}{official website}.} The hybrid model promises to provide a more personalized experience for those who can
come to our institutions in person, while still leaving the course open and accessible to all students from 
around the world to join in remotely to our classrooms.  

\section{Obstacles and challenges}
\label{s.obst}

In organizing and preparing the Python4Physics course, we encountered several challenges, and we believe that future planners of outreach programs might benefit from a checklist of some of these. In this section, we enumerate some of the more significant difficulties we faced and provide suggestions for confronting them. 

\subsection{Advertising and attracting students}

Since many university researchers rarely interact with students outside of the university system, it can be challenging to 
find high school participants. We largely benefited from the experience of staff in our departments and institutions, who were able to connect us with local high school educators. Through these contacts, we were able to either directly or indirectly reach out to high school participants. In several instances, we took the time to go to those high schools and give ``general audience" talks about our lines of research in front of students. As long as it is done in an accessible and enjoyable fashion, this personal touch can go a long way to make students comfortable in coming to your classroom. 

\subsection{Technical resources and \\ accessibility}

Ensuring access to computers and software is of course a challenge for any outreach course that involves programming. 

Given that one of our goals is to make STEM as accessible as possible, we have been mindful that our class requires the students to have access to a computer. Within the USA, many High Schools now provide Chrome Books or similar low-cost laptops during the academic year. Depending on the school district, these laptops may or may not be accessible to students during the summer. Ideally, the institutions hosting an outreach activity like Python4Physics can provide these laptops during the summer, but depending on the school district, this may not be possible.

For our first summer with in-person students, we used a stockpile of notebook computers that we previously used in the university for lab classes. This worked to a degree, but we encountered some difficulties with installing the necessary software, and a fraction of the computers failed to function for the full course. 

This is an ongoing challenge for which we have not yet found an ideal solution. Ideally, external private investors could be sought to provide funds to ensure access to laptops.\footnote{In preparation for the first hybrid Python4Physics course to be hosted in the summer of 2024, we have asked our participants if they have access to a computer in our \href{https://docs.google.com/forms/d/e/1FAIpQLSc6KvknrKx80AP1bggk34KyCrCTEpRHMBhsn9VNQYYjRzmCWA/viewform}{registration page}. }

\subsection{Working with minors}

Because our target audience is comprised of high school students, most participants are under 18 years of age, and this comes with a number of important legal considerations. Our experience has been that institutions like Old Dominion University and the University of California, Berkeley, have clear guidelines and best practices that one must follow when hosting an in-person event. In our case, everyone who was to interact with the students was required to undergo a training and criminal background check. This included our undergraduate and graduate assistants. To avoid delays, it is important to start this process early. Depending on the institution, there may be other additional restrictions, e.g. two or more adults that have gone to said training/checks may need to be present with students at all times. 

Other requirements for working with minors may not be as obvious to those who are more familiar with a college campus environment. For example, depending on the locality, it is prohibited to include minors in a \emph{cc} on a mass email. For sending out course information, therefore, it is important to use \emph{bcc} as a default. We suggest communicating with your local institution or institutions very early about its own guidelines for interacting with minors so that preparations can be made. 

When it came to working with minors remotely, at the beginning of the pandemic we found the guidelines to be less clear. As a result, in collaboration with the \href{https://physics.berkeley.edu/visiting-students/reyes-remote-experience-young-engineers-and-scientists}{REYES} team, we opted to follow the safest possible option. We hosted all classes via \href{https://www.zoom.com/en/products/webinars/?creative=671275477599&keyword=zoom%20webinar&matchtype=e&network=g&device=c&zcid=13537&gad_source=1&gclid=Cj0KCQjwqpSwBhClARIsADlZ_Tk-JI7Hon2Bh1xNpMHv2pRtB6S4bOIsD_1xhkg4sEJcZe7ev9LyyYkaAnuoEALw_wcB}{Zoom Webinar}. Zoom Webinar allows one to have two distinct environments where participants can be placed, e.g. ``\emph{Panel}"
 or ``\emph{Audience}". Depending on where the participant is placed, they may be restricted in how they interact with others. This allowed us to ensure that nobody could communicate directly with students, and it allowed us to control who may activate their cameras or communicate with others. This provided a remarkably safe experience for students, and it made it impossible for anybody to ``\emph{Zoombomb}'' others with any profanity or unwanted behavior.\footnote{
Note, that other companies have similar Webinar platforms that may be more affordable, e.g. \href{https://www.bluejeans.com/nb/node/60106}{BlueJeans}.}

\subsection{Time sink for hosts}

It is inevitable that a program of the scope of Python4Physics 
will take time away from faculty who are already spread quite thin. 
We find that the major time sinks include the time required in developing the material, hosting the actual classes, and miscellaneous administrative burdens.

Although some time costs are unavoidable, we are happy to provide the following tips and resources to help reduce the burden:

\begin{itemize}
\item {\bf Developing material}: We are making all of the material publicly available. This also includes our code that solves numerical tasks given to the students. These code solutions are also shared for the students to self-evaluate their solutions. You are welcome to copy and edit it as you wish. Please find the materials at our project website \href{https://tddyrogers.github.io/python4physics.github.io/}{here}. \\
\item {\bf Hosting class}: We have found that co-hosting classes reduces the burden on each person substantially. In the last couple of years, we have opted to create teams of volunteer faculty/staff, postdocs and graduate students to help host the classes. Depending on the number of lecturers, each person may have to lecture as little as one or two classes. \\
\item {\bf Administrative burden}: Although we cannot remove all administrative burden from those of you interested in hosting this, we are happy to coordinate with you one of your efforts to host the class and provide as much service as possible. An example of how we may be able to help is by sharing our registration pages. Other examples can be discussed in a case-by-case basis. 
\end{itemize}

Although here we list the burdens associated with the course, we would like to emphasize that the experience of working with aspiring young scientists through Python4Physics has been exclusively positive and fulfilling for the instructors.

\section{Lessons and other suggestions}
\label{s.lessons}

Observations we have made while conducting the Python4Physics program have led us to create our own set of guidelines that we feel help us to provide the most fulfilling experience possible for students. While these are only based on our personal experiences, and others may find better ways to achieve their outreach goals, we hope that an account of our own experiences may be useful. 

First, we found that the computational and programming aspects are particularly intimidating to many students, and special efforts to make the programming itself less formidable are always an improvement. Furthermore, we found that we had spent excessive amounts of time on installing software in the earliest versions of the course. It is our experience that any measures that reduce the discussion of specific coding or software tasks will improve the quality of the experience for students. While \emph{exposure} to computers and programming is important, it is preferable to use the limited time and attention available to cultivate broader scientific interests than to focus on technical details. \emph{By motivating and helping students gain some confidence, they will be more likely to dedicate the extra time needed to learn the technical details needed in due time. \bf{Think of yourself more like a motivational coach, rather than a university professor.}}

The need for specialized software (compilers, interpreters, etc) can be minimized by using online coding platforms, so that access to a web browser is all that is necessary. For teaching programming, we typically utilize a single short lecture to explain the basic structure of a generic program. There, we briefly introduce concepts like \emph{if} statements and \emph{for} loops, but with very little detail. An effective way to ensure that most students may take part in some amount of programming without being stuck or left behind is to provide them with nearly completed codes, with only one or a few pieces that need to be finished. Fully completed codes are also provided so that they may check their own work. This provides the majority of students with an experience of ``coding'' in the very short time that is typically available. 

These remarks concerning computer programming also apply to mathematical reasoning more generally. Whenever equations need to be used for a task, the exact final form that is necessary should be provided, in a clear and easy-to-use format, without the need for any manipulations. 

We find that student experiences always seem to improve when we maximize the humor, drama, and visualization that we inject into the specific projects. Current events or popular topics in online discourse provide valuable source material for this. For example, we assigned projects to model epidemics, and we taught some basics of cryptography, again providing examples and sample codes. We provided guidance for visualizing interesting shapes like the Fibonacci spiral. Many resources for fun and whimsical computer projects already exist, and it is worthwhile to consider these.

Students who join the program come with a very wide range of backgrounds in programming, mathematics, and science. By organizing the course as described above, students are able to tailor the experience to their individual levels. Those seeking a greater challenge may choose to build programs that tackle of their own to tackle our assigned projects from scratch, while those encountering the topics for the first time have a framework to build upon. We also encourage a healthy collaborative environment. Among the ways we do this, includes asking more experienced students to help those are coding for the very first time.

\section{Summary}

Our goal is for all students to leave with experiences they can use as starting points for future scientific endeavors. Each year we ask students to provide feedback of their experience. Each year, we get that no less than 70\% ``\emph{felt more confident and enthusiastic in pursuing a career in STEM and conducting research}.

In summary, we strongly believe that Python4Physics is a powerful method to inspire the next generation of scientists. The class requires a minimal amount of funds, and it is easily reproducible. We are sharing our material and experience, with the hopes that others may build from it. 

\section*{Acknowledgments}
RAB was supported in part by the U.S. Department of Energy, Office of Science, Office of Nuclear
Physics under Awards No. DE-AC02-05CH11231.
TCR was supported by the U.S. Department of Energy, Office of Science, Office of Nuclear Physics, under Award Number DE-SC0024715. This work
was also supported by the DOE Contract No. DE- AC05-06OR23177, under which 
Jefferson Science Associates, LLC operates Jefferson Lab. 




\end{document}